\documentclass{article}
\usepackage[utf8]{inputenc}
\usepackage{epsfig}
\usepackage{authblk}
\usepackage{xcolor}
\usepackage{overpic}
\usepackage{amssymb}
\usepackage{amsmath}

\usepackage[margin=1in]{geometry}
\usepackage{multirow}
\usepackage{soul}
\usepackage[colorlinks=true, linkcolor=blue, filecolor=blue, citecolor=blue]{hyperref}
\usepackage[square, comma, sort&compress]{natbib}
\usepackage{lineno}
\title{Challenges of learning multi-scale dynamics with AI weather models: Implications for stability and one solution} 

\author{Ashesh Chattopadhyay$^{1}$, Y.Qiang Sun$^{2}$, and Pedram Hassanzadeh$^{2}$\thanks{pedramh@uchicago.edu}}
\affil{\normalsize $^1$Department of Applied Mathematics, University of California, Santa Cruz, Santa Cruz, 95064, CA}
\affil{\normalsize $^2$Department of Geophysical Sciences, University of Chicago, Chicago, IL 60637}

\date{}

\begin{document}
\maketitle

\section{Abstract}
Long-term stability and physical consistency are critical properties for AI-based weather models if they are going to be used for subseasonal-to-seasonal forecasts or beyond, e.g., climate change projection. However, current AI-based weather models can only provide short-term forecasts accurately since they become unstable or physically inconsistent when time-integrated beyond a few weeks or a few months. Either they exhibit numerical blow-up or hallucinate  unrealistic dynamics of the atmospheric variables, akin to the current class of autoregressive large language models. The cause of the instabilities is unknown, and the methods that are used to improve their stability horizons are ad-hoc and lack rigorous theory. In this paper, we reveal that the universal causal mechanism for these instabilities in any turbulent flow is due to \textit{spectral bias} wherein, \textit{any} deep learning architecture is biased to learn only the large-scale dynamics and ignores the small scales completely. We further elucidate how turbulence physics and the absence of convergence in deep learning-based time-integrators amplify this bias, leading to unstable error propagation. Finally, using the quasi-geostrophic flow and European Center for Medium-Range Weather Forecasting (ECMWF) Reanalysis data as test cases, we bridge the gap between deep learning theory and  numerical analysis to propose one mitigative solution to such unphysical behavior. We develop long-term physically-consistent data-driven models for the climate system and demonstrate accurate short-term forecasts, and hundreds of years of time-integration with accurate mean and variability.






\section{Introduction}
Data-driven models for certain components of the Earth system have emerged to be competitive with physics-based numerical models, e.g., for weather prediction \citep{weyn2020improving,rasp_2020_resnet,schultz2021can,weyn2021sub,chantry2021opportunities,pathak2022fourcastnet,bi2022pangu,lam2022graphcast}, ocean modeling \citep{zeng2015predictability}, sea-ice modeling \citep{andersson2021seasonal,chattopadhyay2023oceannet}, land process modeling \citep{reichstein2019deep}, etc., at a fraction of the computational cost of numerical models. 
In the context of atmospheric modeling, AI-based weather models have emerged to be more accurate than state-of-the-art numerical weather predictions models for short-term forecasts~\citep{lam2022graphcast,bi2022pangu}. However, the physical consistency of these models suffer when integrated for long time scales, beyond a few weeks or months, and produce either numerical instabilities \citep{scher2019weather,weyn2020improving,chattopadhyay2021towards,pathak2022fourcastnet,keisler2022forecasting} or numerically bounded unrealsitic hallucinations. An example of such a hallucination is in the form of excessive diffusion~\citep{bonavita2023limitations} and eventual saturation, as shown in Keisler \textit{et al.}~\citep{keisler2022forecasting}, where the graph neural network-based weather model showed a saturated atmosphere after $100$ days of integration. Other examples of such hallucinations are unphysical drifts in the dynamical variables, e.g., in Scher \textit{et al.} \citep{scher2019weather}, long-term integration showed an unphysical reversal in wind direction. State-of-the-art AI-based weather models such as Pangu3D~\citep{bi2022pangu} and GraphCast~\citep{lam2022graphcast} have similar unphysical drifts although detecting them is nuanced and would be discussed in this paper. Similar unphysical drifts can be seen in the context of ocean modeling as well~\citep{agarwal2021comparison}. While the instantaneous snapshots of the dynamical variables after long-term integration with AI-based weather models may not always reveal a persistent drift in the dynamics, time-averaged quantities for these variables reveal the physical inconsistencies. Such unphysical features of AI-based weather models limits their usefulness and reduces their trustworthiness, especially for predicting extreme events and influencing real-time operational decision-making. The physical consistency of data-driven weather models beyond $5-12$ weeks would allow improved sub-seasonal-to-seasonal-scale probabilistic predictions of the atmosphere \citep{weyn2021sub}, which is currently a grand challenge in the Earth sciences \citep{mariotti2018progress}. 

More recently, the Allen Climate Emulator (ACE), has shown robust hundreds of years of climate emulation~\cite{watt2023ace,duncan2024application} when trained on state-of-the-art climate model outputs at 100Km resolution (much coarser than the 25Km resolution of reanalysis data on which other AI-based weather models are trained on). However, ACE requires the \textit{true} boundary conditions from the ocean, i.e., sea-surface temperature (SST) as well as the top-of-the-atmosphere (TOA) radiation to be incorporated from an already available climate model run. Hence, ACE does not data-drivenly evolve the boundary conditions, but rather \textit{force} the boundary conditions and hence the seasonal cycle of the climate system on the data-driven emulated trajectory. This approach has been successful in constraining the emulated trajectory to maintain stability and physical consistency but is dependent on being able to generate the boundary conditions from an external dataset (in this case a climate model run). A similar approach was adopted by~\cite{subel2024building} to generate stable ocean emulation as well with a fully data-driven model. The physical reason for such stability arising due to the imposition of boundary conditions has not been studied  

Longer-term integration at the time-scale of hundreds of years, would allow us to obtain a large number of ensemble members of climate simulations that would enable long lead-time prediction of the distribution of extreme events, e.g., heatwaves \citep{mckinnon2016long,chattopadhyay2019analog} and couple these models to other components of numerical Earth system models for cost-efficient Earth system predictions. Earlier work on generating large ensembles of emulated trajectories for data assimilation has been shown to be successful in Chattopadhyay \textit{et al.}~\cite{chattopadhyay2020deep,chattopadhyay2023deep} with physics-constrained deep learning models. Recent work on generative model-based emulation of large ensembles of prediction for weather prediction has shown state-of-the-art performance against ensemble-based numerical weather prediction models~\cite{price2023gencast}. However, all of these works have focused on shorter time scales ($\le 15$ days) since these deep learning models are not consistent for longer time scales. 

\textit{To be useful for long-term emulation, it is not sufficient for an AI-based weather model to produce bounded predictions, but also physically consistent ones, with correct long-term mean, Fourier spectrum, and variability.} While some of the current AI-based weather models such as the spherical Fourier neural operator~\citep{bonev2023spherical} and Pangu3D~\citep{bi2022pangu} produces bounded and \textit{``seemingly realsitic"} instantaneous snapshots of the dynamical variables at long integration time scales, that is often misleading. Time-averaged quantities such as mean would reveal that these models have generated unphysical dynamics during long-term emulation. 


Moreover, while some metrics in the short-term skills of these models may have high accuracy, e.g., root-mean-squared-error (RMSE) and anomaly correlation coefficient (ACC), it does not indicate that the model would be long-term stable and yield physically-consistent emulation with accurate mean and variability. The cause of this instability or unphysical drift in data-driven models of  weather is largely unknown and hence the remedies, in the form of energy/enstrophy constraints, training with adversarial noise \citep{suadversarial}, etc.,  are often ad-hoc and can only delay the onset of instabilities/unphysical traits without actually removing them.

In this paper, for the first time, we reveal the fundamental cause of this hallucination through the lenses of deep learning theory and nonlinear physics. We show that the unphysical features are a result of a fundamental inductive bias called \textit{spectral bias}, that exists in all deep learning architectures, which is further amplified during autoregressive prediction due to error propagation through non-convergent integration schemes. This combination of spectral bias and error propagation would affect \textit{all} data-driven models that are used to predict \textit{any} turbulent flow in engineering or natural sciences \citep{li2021markov}. Finally, we propose a physics-inspired, architecture-agnostic framework, \textbf{FOU}rier-\textbf{R}unge-\textbf{K}utta-with-\textbf{S}elf-supervision, FouRKS, to ensure physical consistency. The key contributions in this paper are as follows:
\begin{itemize}
    \item Delineating the role of spectral bias and error propagation in the onset of unphysical features for any data-driven model of turbulent flows.

    \item A Fourier-based regularization strategy to mitigate spectral bias during the training of a model.

    \item A convergent higher-order time integrator implemented as a custom differentiable layer inside the model during training.

    \item A self-supervised spectrum correction strategy to ensure physical consistency during autoregressive prediction.
\end{itemize}

We demonstrate FouRKS' effectiveness to produce long-term stable and physically-consistent climate simulations for hundreds of years, with accurate mean, spectrum, and variability (for QG), using the two-layer quasigeostrophic (QG) and ECMWF Reanalysis 5 (ERA5) dataset.  

\section{Results}\label{sec2}
In this section, we would first show how state-of-the-art AI-based weather models show instability or physical inconsistencies under long-term integration, when time-averaged quantities are investigated. We would then describe a universal cause for such physical inconsistencies, i.e., spectral bias, and then provide a mitigative solution to ensure that the data-driven models produce physically consistent emulation for \textit{any nonlinear multi-scale system}. We will demonstrate our results using the QG system and ERA5 data.

\subsection{A universal cause: Spectral bias}
\begin{figure*}[ht]
  \centering
\includegraphics[width =0.95\textwidth, trim={0cm 0cm 0cm 0cm},clip]{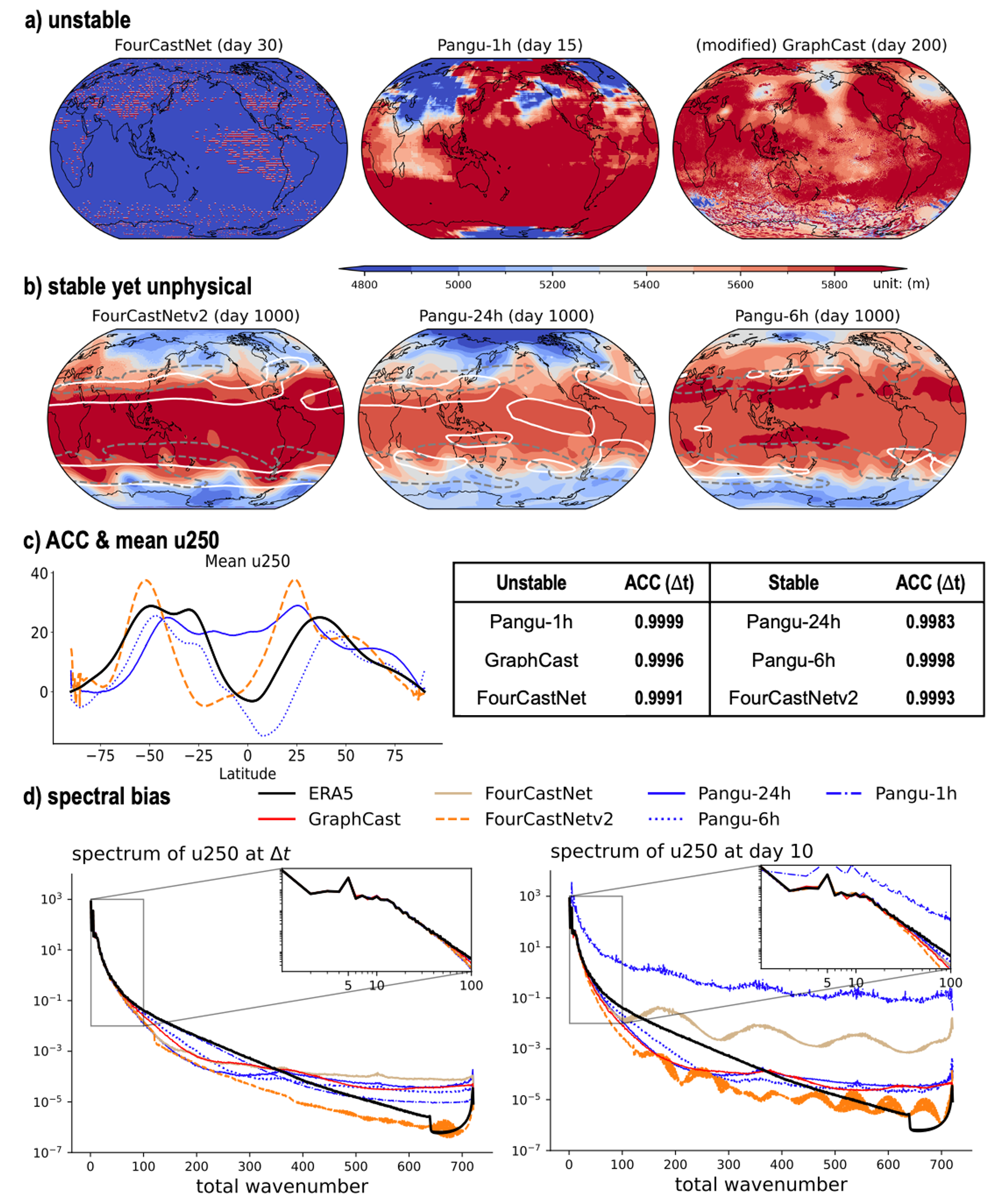}
  \caption{
  Hallucinations in Pangu3D~\citep{bi2022pangu}, GraphCast~\citep{lam2022graphcast}, FourCastNet~\citep{pathak2022fourcastnet}, and FourCastNetv2~\citep{bonev2023spherical} trained on $0.25^{\circ}$ ERA5 data shown in the Z500 and U250 field. (a) Snapshots of FourCastNet, Pangu3D with 1h time step, and autoregressive GraphCast shows unstable blow-up after a few days. (b) FourCastNetv2, Pangu3D with $24$h time step, and Pangu3D with $6$h time step remains stable but shows unphysical characteristics better shown by (c). (c) To understand the unphysical characteristics in (b), we show long-term mean U250 of all the models that do not match the true U250 mean from ERA5 data. (d) Spectral bias showing that the spherical harmonic-based Fourier spectrum of predicted U250 even for the first time step of prediction does not match the true spectrum of ERA5 at the first time step of prediction although ACC shown in (c) is $\approx 1$. (e) Spectral bias growing by $10$ days of prediction.}
  \label{fig:pangu}
\end{figure*}

We first demonstrate that long-term emulation of state-of-the-art AI-based weather models become either unstable or physically-inconsistent. Figure.~\ref{fig:pangu}(a) clearly shows a numerical blow-up of the Z500 fields from FourCastNet~\citep{pathak2022fourcastnet}, Pangu3D with a time step of $1$h, and GraphCast. Here, both GraphCast and FourCastNet used a time step of $6$h. FourCastNetv2 and Pangu3D with time steps of $6$ and $24$h do not blow up and may ``appear to be realistic". However, they become physically inconsistent. In order to gauge the inconsistency, instead of investigating instantenous snapshots of the dynamics predicted by the models at long time scales, we should compute long-term statistics, such as the zonal- and time-mean of the dynamics averaged over several years of emulation. We show that zonal- and time-mean Z500 and U250 produced by Pangu3D with time step of 24h and 6h and FourCastNetv2 with time step of $6$h averaged over 10 years of emulation in Fig.~\ref{fig:pangu}(c). We can see that the mean for either Z500 and U250 do not correspond to the mean computed from ERA5 data. Pangu3D shows unphysically large U250 mean value near the equator wherein the observed mean U250 should be close to $0$. Moreover, FourCastNetv2 shows unphysical variations of both mean Z500 and U250 near the poles of the Earth. Hence, the instantaneous values of the dynamical variables from AI-based weather models can be misleading when gauging the quality of the models for long-term emulation. Figure~\ref{fig:pangu}(d) demonstrates spectral bias~\citep{xu2019frequency,rahaman2019spectral,basri2020frequency} in these models from the first time step of prediction ($6$h and $24$h for Pangu3D, and $6$h for GraphCast and FourCastNetv2). Here, the Fourier spectrum of the predicted variable does not match the true spectrum of the variable (obtained from ERA5), even when the accuracy of prediction is very high. Spectral bias is the inability of these models to accurately capture the high wavenumbers or small scales in the predicted variables. This is an inherent inductive bias in deep neural networks, wherein, during training, the learnable parameters of the networks fail to update, once the low-wavenumber components of the true signal's Fourier spectrum is learned, thereby prohibiting the network from learning the high-wavenumber components. As such, the predicted high-number component may either be above (below) the true spectrum indicating instability (excessive diffusion). However, the common metrics used to determine the skill of the networks for forecasting tasks, such as ACC, are biased by the large scales and do not get affected by the networks' inability to capture the high-wavenumber part of the true spectrum and remain high even when the true and predicted spectrum does not match in the small scales.  Thus, high values of ACC for short-term forecasts  is not a reliable indicator for the model's physical consistency for long-term emulation. This model error incurred in the small scales at the first time step grows nonlinearly and affects the large-scale flow in future time steps. If the large-scale flow shows excessive diffusion then the models remain stable but eventually becomes physically incosistent, while if the large-scale spectrum is overestimated as compared to the true one, the models become unstable. GraphCast, however, appears to have constant spectral bias, and its instability could be connected to some other mechanism that is yet unexplored in data-driven models. Predictions from GenCast~\citep{price2024probabilistic}, which are inherently stochastic in nature, show lower spectral bias, as is evident in Figure 2 panels (g)-(m) in the paper \cite{price2024probabilistic} than most of the models (except for FouRKS) in the this paper. There can be several reasons for the decrease in spectral bias, including the backward integration in a diffusion model that resembles autoregressive integration in wavenumber space~\citep{rissanen2022generative}. GenCast is trained on multiple different resolutions of ERA5 data and fine-tuned on $0.25^{\circ}$. Furthermore, the extent of spectral bias depends on the chosen variables whose Fourier spectrum is being investigated as well. Dynamical variables like Z500 show more spectral bias than, e.g., relative humidity. Finally, visualizing the spectrum in a log-log plot might appear to show lower spectral bias. For consistency, we have also plotted the Fourier spectrum of the AI-weather models in log-log scale at both $0.25^{\circ}$ and $1^{\circ}$ in supplementary Figure S$5$.  

\begin{figure*}[ht]
  \centering
\includegraphics[width = \textwidth, trim={0cm 0cm 0cm 0cm},clip]{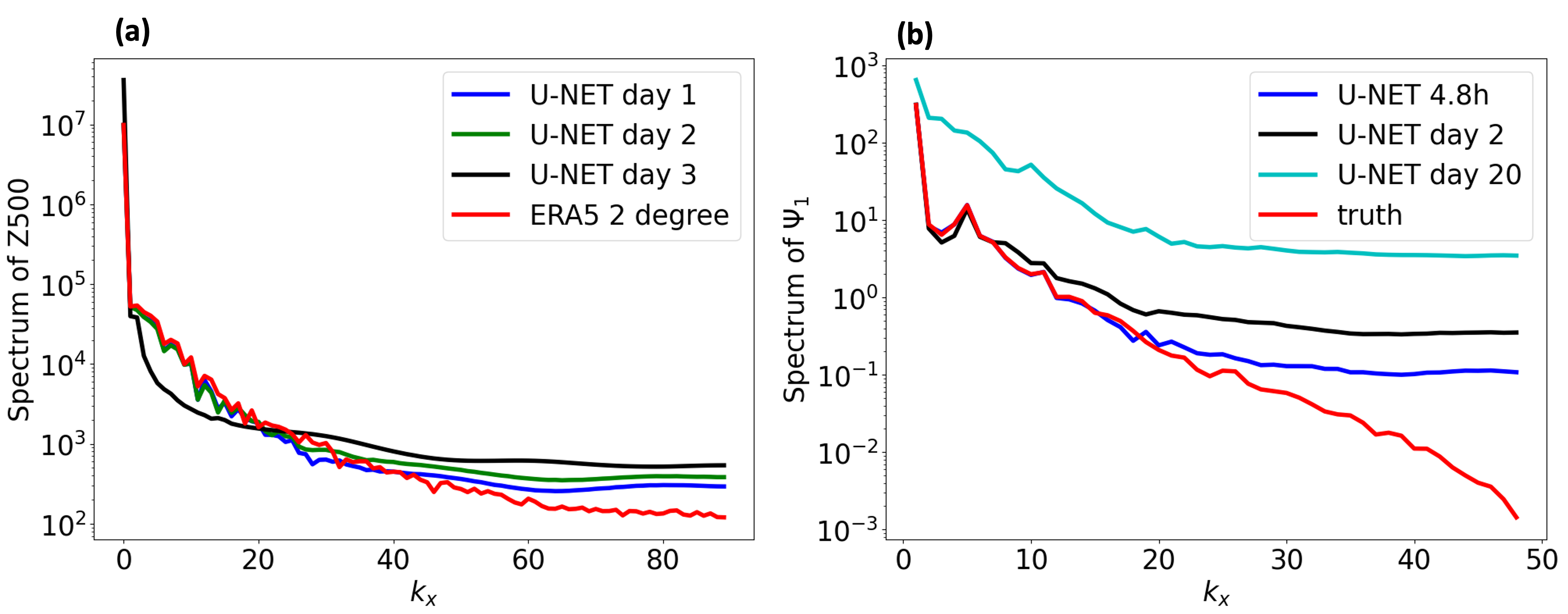}
  \caption{Long-term instabilities in a simple  U-NET-based digital twin (section~\ref{sec:Unet}) trained on $2^{\circ}$ ERA5 data (section~\ref{sec:ERA5}) and QG simulations (section~\ref{sec:QG}). (a) Latitude-averaged instantaneous Fourier spectrum of predicted Z500 fails to capture the small-scale part of the true spectrum beyond $k_x \geq 25$. (b) Latitude-averaged instantaneous Fourier spectrum of predicted $\psi_1$ with U-NET shows that even for QG simulations, the small-scale part of the true spectrum cannot be captured right from the first time step of prediction ($4.8$ hrs).}
  \label{fig:unstable snapshots}
\end{figure*}

To ensure that spectral bias in deep learning and the subsequent error propagation is not an isolated phenomenon in complex vision transformers, neural operators, or multi-level graph architectures, we demonstrate spectral bias with a simple U-NET on a 2-layer quasigeostrophic model and coarse-resolution $2^{\circ}$ ERA5 data with Z500, Z50, and Z850 as the only variables in Fig.~\ref{fig:unstable snapshots}.  We show that even a simple U-NET trained on either $2^{\circ}$ ERA5 or QG data yields accurate short-term predictions but becomes unstable and shows unphysical values of Z500 (in case of ERA5) and upper- and lower-level stream function, $\psi_1$ and $\psi_2$ (in case of QG, where $1$ and $2$ are for the upper and lower level) after $100$ days of autoregressive emulation. The zonal Fourier spectrum of the predictions, as shown in Fig.~\ref{fig:unstable snapshots}(a)-(b), reveal that U-NET trained on $2^{\circ}$ ERA5 data, and U-NET trained on QG data, fail to capture the small-scale part of the true Fourier spectrum of the fields (Z500 in case of ERA5 and $\psi_1$ in case of QG) during autoregressive prediction, starting from the first time step. The zonal Fourier spectrum has been computed since both the QG system and ERA5 have a jet moving along the zonal direction. 

Spectral bias in deep neural networks manifests itself as an epistemic error. During the time of autoregressive prediction, the nonlinear interaction between the small and large scales \citep{cambon1999linear} of the system, amplifies the errors in the predicted large-scale part of the true spectrum as well. Autoregressive emulation at long time scales ($\approx 100-200$ days) with such epistemic errors would finally result in an unphysical value of the emulated field. For example, in Fig.~\ref{fig:QG_stable}(a) and (b), we show the time- and zonal-mean structure of the emulated upper- and lower-level velocity respectively, $\left<\overline{u}_1\right>$ and  $\left<\overline{u}_2\right>$ (blue dashed line). It has no resemblance to the true time- and zonal-mean structure. For QG, the model emulation shows spectral bias in predicted lower-level stream function, $\psi_2$, as well (Fig.~S1(a)), which results in unphysically large values of the meridional heat flux during autoregressive emulation as shown in Fig.~S1(b). This leads to an increase in the momentum flux as shown in Fig.~S1(c), finally resulting in an unphysical value of lower-level velocity, $\left<\overline{u}_2\right>$ (Fig.~S1(d)). 

It must be highlighted here, that all deep neural networks, including fully-connected \citep{rahaman2019spectral}, convolutional (this paper), operator (this paper), transformer-based (this paper), and generative \citep{schwarz2021frequency} suffer from this epistemic error in the form of spectral bias, wherein they fail to learn the high-wavenumber components of the signal that they are trained to predict. This error is an inductive bias, which means that one cannot mitigate it by training on more data, longer epochs, or having more capacity in the network. However, in most classification problems involving natural images, capturing the low-wavenumber components of the signal is often enough to predict the probability of the correct class. For problems in multi-scale partial differential equations (PDEs) with tightly coupled spatiotemporal scales, such as in atmospheric, oceanic, and engineering turbulence, such epistemic errors along with their non-trivial and nonlinear propagation and subsequent amplification in the large scales would render all data-driven emulators useless. 

Furthermore, one should note that while spectral bias is a universal and fundamental cause of instability or halluciantions, there may be other contributing causes as well, e.g., unphysical accumulation of kinetic energy near the poles (due to geometric distortion of data) in data-driven models of the atmosphere may lead to unphysical drifts as well. This has been addressed in~\cite{weyn2020improving} through better geometric representations and through the spherical Fourier neural operators in~\cite{bonev2023spherical}. Eliminating any of such many causes would not mitigate spectral bias. To elucidate that, we have also considered QG as an example where the distorting effect of poles is absent and yet the emulations become unstable due to spectral bias.   
  
\subsection{Mathematical demonstration of spectral bias in a single layered neural network}

To demonstrate the emergence of spectral bias mathematically, we consider a single layer fully-connected neural network with 1D scalar input, $u$, 1D analytical function as a label, $f(u)$, and activation function $tanh(u)$. We further assume that the hidden layer has $m$ nodes; the output of the network, $h(u)$, can be written as:

\begin{eqnarray}
    h(u)=\sum_{j=0}^{j=m} v_j\sigma\left(w_j u+b_j\right),
    \label{eq:basic_NN}
\end{eqnarray}

where $v_j$ are the weight elements that connects to the ouput, $w_j$ are the weights for the hidden layer, $\sigma$ represents the $tanh$ activation function for compactness. 

Here, we note that the Fourier transform of $\sigma$ is given by:

\begin{eqnarray}
    \hat{\sigma}(k)=\frac{-iuk}{sinh\left( \pi k/2\right)},
    \label{eq:tanh_fft}
\end{eqnarray}

where, $k$ is the wavenumber and $i$ is the imaginary number. 

\begin{eqnarray}
    \hat{h}\left(k\right) = \sum_{j=0}^{j=m}\frac{2 \pi v_j iu}{|w_j|} exp \left (\frac{iu b_jk}{w_j}\right) \frac{1}{exp\left( \frac{-\pi k}{2w_j}\right)-exp\left( \frac{\pi k}{2w_j}\right)}.
\end{eqnarray}

We define the Fourier transform of the deviation between the Fourier transform of the neural net's output and Fourier transform of the label as $\hat{D}(k)$:
\begin{eqnarray}
    \hat{D}(k) = \hat{h}(k)-\hat{f}(k).
\end{eqnarray}

We can write, $\hat{D}(k)$ as $A(k)e^{\phi(k)}$, where $A$ is the amplitude and $\phi$ is the phase of the deviation. The loss function in Fourier space, $\hat{L}(k)$ at wavenumber, $k$, is given as:
\begin{eqnarray}
    \hat{L}(k)= \frac{|\hat{D}(k)|^2}{2},
\end{eqnarray}
and the total loss, $L$, is give by:
\begin{eqnarray}
    L= \int_{-\infty}^{\infty}\hat{L}(k)dk.
\end{eqnarray}

We can now compute the partial derivatives of $\hat{L}(k)$ with respect to $v_j$, $w_j$, and $b_j$. They are given by:

\begin{eqnarray}
    \frac{\partial \hat{L}(k)}{\partial a_j} = \frac{2 \pi}{w_j} sin\left( \frac{b_jk}{w_j} - \phi(k) \right)\mathcal{E}_0,
\end{eqnarray}

\begin{eqnarray}
    \frac{\partial \hat{L}(k)}{\partial b_j} = \frac{2 \pi a_j b_jk}{w^2_j} cos\left( \frac{b_jk}{w_j} - \phi(k) \right)\mathcal{E}_0,
\end{eqnarray}

\begin{align}
\begin{split}
    \frac{\partial \hat{L}(k)}{\partial w_j} &= sin\left( \frac{b_jk}{w_j} - \phi(k) \right)\left( \frac{\pi^2a_j k}{w_j^3}\mathcal{E}_1 - \frac{2\pi a_j}{w^2_j}\right)\mathcal{E}_0 - \\
    &\frac{2\pi a_j b_j k}{w^3_j}cos\left( \frac{b_jk}{w_j} - \phi(k) \right)\mathcal{E}_0.
    \end{split} 
\end{align}

Here, $\mathcal{E}_0$ and $\mathcal{E}_1$ are given by:

\begin{eqnarray}
    \mathcal{E}_0 = \frac{sgn(w_j)A(k)}{exp\left( \frac{-\pi k}{2w_j}\right)-exp\left( \frac{\pi k}{2w_j}\right)},
\end{eqnarray}

and

\begin{eqnarray}
    \mathcal{E}_1 = \frac{exp\left( \frac{-\pi k}{2w_j}\right)+exp\left( \frac{\pi k}{2w_j}\right)}{exp\left( \frac{-\pi k}{2w_j}\right)-exp\left( \frac{\pi k}{2w_j}\right)}.
\end{eqnarray}

\begin{figure*}[ht]
  \centering
\includegraphics[width =0.95\textwidth, trim={0cm 0cm 0cm 0cm},clip]{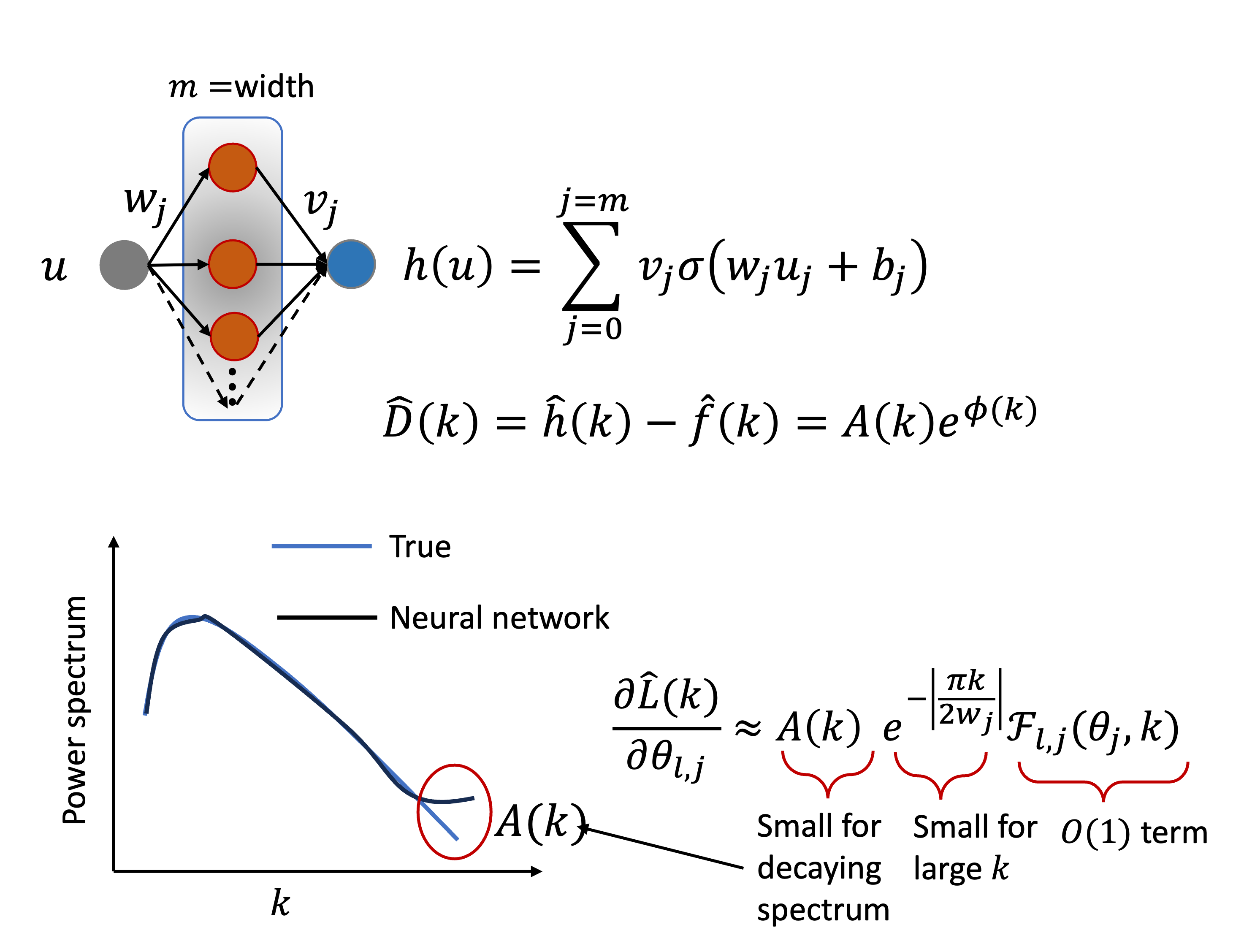}
  \caption{
  Toy 1D example of a single layered neural network fitting a scalar-valued function $f(u)$ as a function of scalar, $u$. As learning progresses, the total gradient of $\hat{L}(k)$ as a function of parameters, $\theta_j$ becomes smaller since $A(k)$ becomes smaller at small wavenumbers, $k$. However, for large wavenumbers, $k$, where $A(k)$ is nonzero, the decaying spectrum of turbulence, ensures that the values is small; moreover the exponential multiplicand, $exp(-|\pi k/2w_j|)$ ensures that the total gradient of $\hat{L}(k)$ remains small and the parameters, $\theta_j$ are not updated, leading to a bias in the small scales, i.e, large values of $k$.}
  \label{fig:SB_schematic_figures}
  \end{figure*}

The gradient of $\hat{L}(k)$ with respect to all the parameters, written compactly as $\theta_j=\left[w_j, b_j,  a_j\right]^T$ is given by:
\begin{eqnarray}
    \frac{\partial \hat{L}(k)}{\partial \theta_{l,j} } \approx A(k)exp\left(-|\pi k/2w_j|\right)\mathcal{F}_{l,j}\left(\theta_j,k\right),
    \label{eq:Sb_eq}
\end{eqnarray}

where $l=1$ corresponds to $w_j$, $l=2$ corresponds to $b_j$, and $l=3$ corresponds to $a_j$. $\mathcal{F}_{l,j}\left(\theta_j,k\right)$ is an $O(1)$ term (see~\cite{xu2019frequency} for details). Focusing our attention on Eq.~\ref{eq:Sb_eq}, we can deduce that, during the early phase of training when $\hat{h}(k)$ and $\hat{f}(k)$ are sufficiently different, $A(k)$ is large for all values of $k$, and hence $\frac{\partial \hat{L}(k)}{\partial \theta_{l,j} }$ is large which leads to the updates of parameters $\theta_{j}$ and the neural network learns. As the networks learns the large scales, the value of $A(k)$ becomes close to $0$ for small values of $k$ as shown in the schematic given in Fig.~\ref{fig:SB_schematic_figures}. However, for large values of $k$, although $A(k)$ is nonzero, the decaying nature of the power spectrum ensures that the values of both $\hat{f}(k)$ and $\hat{h}(k)$ are small, and hence, $A(k)$ is \textit{small}. For example, the ratio of the power between the smallest and largest $k$ in the QG system is $\approx 10^{6}$. Furthermore, this $A(k)$ is multipled by $exp(-|\pi k/2w_j|)$ which decreases with an increase in $k$. The product of these two small quantities renders the value of the gradient, $\frac{\partial \hat{L}(k)}{\partial \theta_{l,j} }$ to be very small. Hence, the parameters, $\theta_{j}$ will practically stop updating and create this bias, ``spectral bias" in the small scales, i.e., large values of $k$. While the analysis here, has been conducted for a simple single layered neural network with one hidden layer and $tanh$ activations, this can be extended to ReLu networks as well with deeper layers, although several infinite width approximations would be required to derive an analytical form of the gradient, $\frac{\partial \hat{L}(k)}{\partial \theta_{l,j} }$, in such a case.~\cite{xu2019frequency} conducts a comprehensive analysis of such ReLU networks for 1D functions. The analysis in the section has also been reiterated from~\cite{xu2019frequency}. 

\subsection{A solution: FouRKS (FOUrier-Runge-Kutta-with-Self-supervision) }
In order to ensure long-term physical consistency of AI-based weather models, we have adopted a principled approach to design FouRKS (shown in Fig~\ref{fig:fourks_schematic}),  an architecture-agnostic framework, the details of which are described in section~\ref{sec:fourks_method}. FouRKS addresses spectral bias via a novel spectral regularizer (section~\ref{sec:spec_regul} that penalizes the high wavenumbers during training, reduces error growth via a $4^{th}$-order Runge-Kutta (RK4) integrator implemented as a custom differentiable layer during training, and finally, a self-supervised spectrum-correction strategy to ensure that errors in the small scales are not allowed to propagate into the large scales during autoregressive prediction (inference stage). FouRKS can be used with any deep learning architecture, where the deep learning model is used to predict the residue of the underlying partial differential equation (PDE) of the system (more details in section~\ref{sec:fourks_method}).

\subsubsection{Performance on QG}
Applying FouRKS with a U-NET, on the QG system, we demonstrate stable and physically consistent long-term autoregressive emulation for $300000$ days for the first time. Figure~\ref{fig:QG_stable} shows the long-term stable statistics obtained using the emulation from FouRKS after $300000$ days of autoregressive predictions on the QG system. Fig.~\ref{fig:QG_stable}(a)-(b) show that FouRKS has accurately captured the time- and zonal-mean structure of the upper- and lower-level velocities, $\left<\overline{u}_1\right>$, and $\left<\overline{u}_2\right>$ respectively. Figure~\ref{fig:QG_stable}(c)-(d) show that FouRKS can accurately predict the structure of the empirical orthogonal functions (EOFs) which demonstrates its ability to capture the internal variability of the QG system. 

\subsubsection{Performance analysis of each component of FouRKS}
Here, we investigate the effect of each component of FouRKS individually and together on short-term prediction skill and long-term stability. 

In Fig.~\ref{fig:ablation_study}(a), the black solid line shows the root mean-squared error (RMSE)  growth of the FouRK framework without the spectral regularizer (i.e., only the effect of RK4 integrator, as described in section~\ref{sec:methods_RK4} as a custom layer can be seen) as a function of time, during autoregressive prediction. As expected from a higher-order integrator, the RK4 layer dampens the growth of error as compared to the baseline U-NET (blue dashed line), for a longer period of time ($\approx$ 30 days). However, by $60$ days, the RMSE grows substantially and renders the predictions unphysical. Contrary to the RK4 layer, the spectral regularizer (section~\ref{sec:spec_regul}) pushes the RMSE error to grow faster than the regular U-NET (Fig.~\ref{fig:ablation_study}(b)). While it is not immediately clear as to why the spectral regularizer alone, has such an effect on the error growth in time, it must be kept in mind that, the regularizer only assists in capturing the small scales in a single time step of prediction. The effect of this regularizer on long-term autoregressive error propagation has not been studied and is not trivial owing to the absence of a theoretical framework for studying error propagation through neural network-based time integrators. Fig.~\ref{fig:ablation_study}(e) and (f) show the effect of the RK4 layer and the spectral regularizer on the Fourier spectrum of the predictions of $\psi_1$ (the effect on $\psi_2$ is similar, and not shown for brevity). While the higher-order integration scheme generally produces a more accurate state prediction whose Fourier spectrum better matches the true spectrum, it still cannot capture the smallest scales. However, the spectral regularizer alone can allow the network to produce predictions whose Fourier spectrum matches the true Fourier spectrum up to the smallest scales (blue solid line in Fig~\ref{fig:ablation_study}(f)) for a single time step. Combining RK4 and the spectral regularizer in FouRK allows the RMSE error to remain stable for $\approx 30$ days while marginally outperforming the baseline U-NET for short-term prediction ($3-5$ days) before eventually increasing to unphysically large values by $\approx  60$ days, as shown in Fig~\ref{fig:ablation_study}(c). This is because, despite the spectral regularizer diminishing spectral bias, and the RK4 integrator dampening error growth, the nature of the interaction of nonlinear scales in turbulence would amplify even the smallest of errors in the high wavenumbers and eventually render the emulation useless. The self-supervised spectrum-correction strategy in FouRKS, described in section~\ref{sec:SSL} is thus, an essential component for long-term stability. FouRKS allows the RMSE error to remain bounded for long-term autoregressive predictions ($\approx$ 200 days), shown in Fig.~\ref{fig:ablation_study}(d). Moreover, with FouRKS, the  Fourier spectrum of the predicted state matches all the scales of the true spectrum, even after $200$ days of prediction (Fig.~\ref{fig:ablation_study}(h)).

\subsection{Performance on ERA5}
Here, for the first time, we show that FouRKS with a U-NET trained on $2^{\circ}$ Z500, Z50, and Z850 variables from ERA5 data can produce stable and physically consistent autoregressive predictions for up to $10$ years. As shown in Fig.~\ref{fig:Z500}(a), the time- and zonal-mean structure of the predicted Z500 matches that of ERA5 while the other AI-based weather models show an unphysical structure. We further investigate the Fourier spectrum of Z500 predicted by FouRKS as compared to ERA5 at $2^{\circ}$ resolution in Fig~\ref{fig:Z500}(b) at the end of $10$ years of emulation. The perfect match between the spectrums indicate that FouRKS can accurately mitigate spectral bias. However, comparing the other AI-based weather models with ERA5 data at $0.25^{\circ}$ resolution show an unphysical Fourier spectrum after $10$ years of emulation. It must be noted here that for the other AI-based weather models shown in Fig~\ref{fig:Z500}, we have only considered the ones that remain numerically stable after $10$ years of emulation. We have not computed the EOFs for Z500 data since FouRKS in this case since there is no seasonal cycle imposed on the FouRKS output. One can easily adopt an approach such as ACE~\citep{watt2023ace} to impose the seasonal cycle here as well; however in that case FouRKS would not remain a fully data-driven model. Other approaches to reliably evolve the boundary conditions for the atmosphere needs with deep learning models need to be investigated. 

\begin{figure*}[ht]
  \centering
\includegraphics[width = \textwidth]{FouRKS.png}
  \caption{Schematics for each component of the FouRKS framework and the baseline U-NET. More details about each of the components can be found in section~\ref{sec:Unet} and section~\ref{sec:fourks_method}.}
  \label{fig:fourks_schematic}
\end{figure*}

\begin{figure*}[ht]
  \centering
\includegraphics[width=\textwidth,angle=0,trim={0cm 0cm 0cm 0cm},clip]{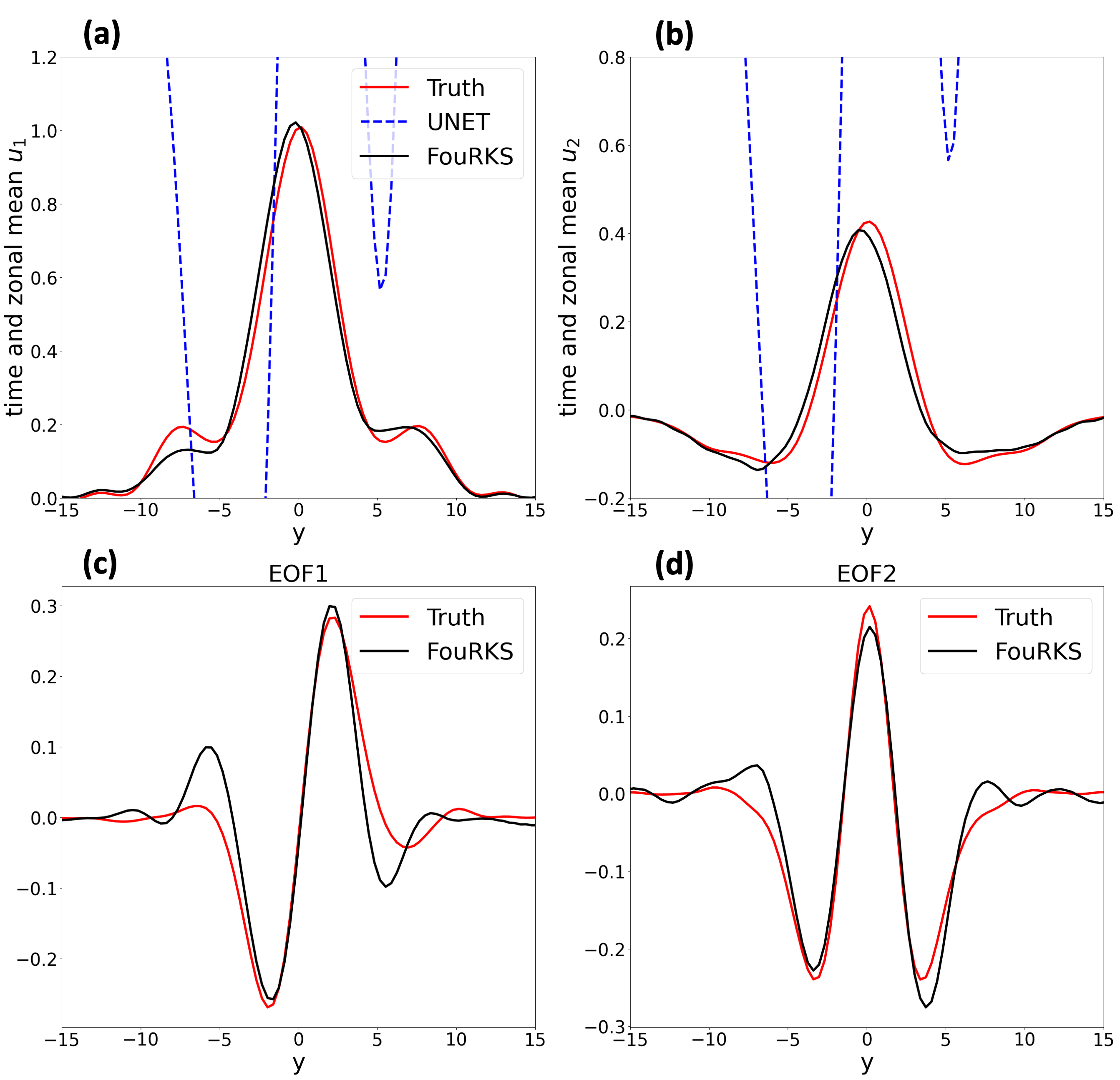}
  \caption{Long-term statistics showing the mean, PDF, and variability of predicted dynamics of QG using FouRKS (section~\ref{sec:fourks_method}) and baseline U-NET (section~\ref{sec:Unet}). The mean, PDF, and EOFs have been computed over $300000$ days of prediction. (a) Zonal- and time-mean of upper-level velocity, $\left<\overline{u}_1\right>$, predicted by FouRKS and true $\left<\overline{u}_1\right>$ shows excellent agreement while U-NET's predicted $\left<\overline{u}_1\right>$ is unphysical. (b) PDF computed with predicted $\psi_1$ with FouRKS shows better agreement with the true PDF as compared to the PDF obtained from the training data. PDF obtained from baseline U-NET could not be plotted with the same axis ranges owing to unphysically large values of the predicted fields. (c) EOF1 from FouRKS shows agreement with the true EOF1. EOF1 from baseline U-NET could not be computed since the predictions from U-NET are unphysically large after $300000$ days, making the numerical computation of EOFs infeasible. (d) Similar to (c) but for EOF2.}
  \label{fig:QG_stable}
\end{figure*}

\begin{figure*}[ht]
  \centering
\includegraphics[width=\textwidth,angle=0,trim={0cm 
0cm 0cm 0cm},clip]{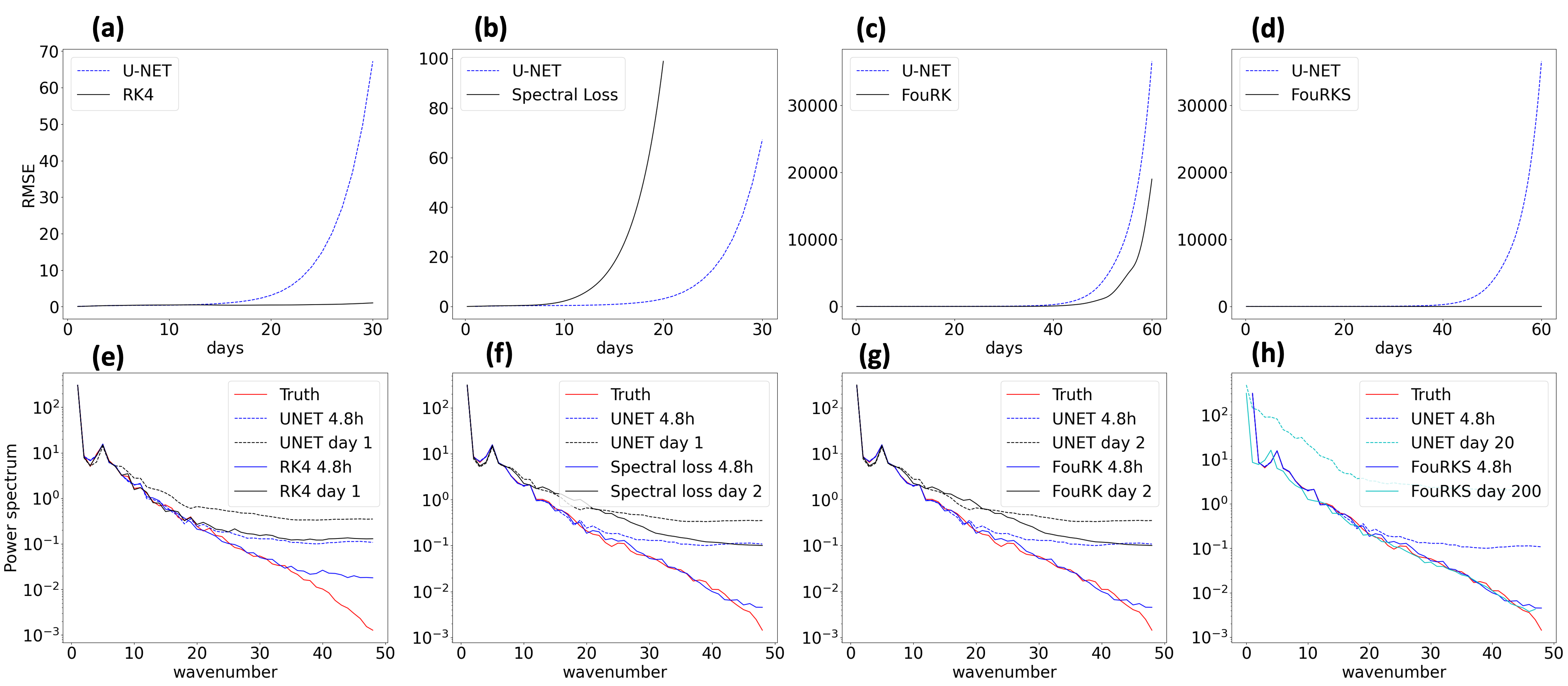}
  \caption{Growth of error in physical and spectral space during autoregressive prediction with different components in FouRKS (section~\ref{sec:fourks_method}). (a) Relative RMSE in FouRK without the spectral regularizer (black solid line) and baseline U-NET (blue dashed line) shows that the RK4 integrator diminishes the error growth during prediction as opposed to the baseline U-NET. It would however keep growing and become unphysical by $\approx 60$ days. (b) Relative RMSE in FouRK without the RK4 integrator, showing that the spectral regularizer does not diminish the error growth at all. (c) Relative RMSE in FouRK shows improvement in error growth as compared to baseline U-NET, but would still become unphysical by $\approx 60$ days. (d) Relative RMSE in FouRKS shows long-term stability as compared to baseline U-NET which becomes unphysical after $200$ days. (e) Fourier spectrum of predicted $\psi_1$ with FouRK ($\hat{\psi}_1$) without the spectral regularizer at the first time step of prediction ($4.8$ hrs) shown with solid blue line and day 1 (solid black line) as compared to U-NET (dashed lines). (f) Fourier spectrum of FouRK's predicted $\psi_1$  without the RK4 integrator. The spectral regularizer alleviates spectral bias by capturing the full Fourier spectrum up to the smallest wavenumbers (solid blue line) as opposed to baseline U-NET. (g) Same as (f) but with FouRK having both the spectral regularizer and the RK4 integrator. (h) Fourier spectrum of predicted $\psi_1$ with FouRKS shows that the spectrum does not curl up during autoregressive prediction, even at $200$ days (solid magenta) as opposed to baseline U-NET whose predictions lose track of even the large scales (dashed magenta).}
  \label{fig:ablation_study}
\end{figure*}

\begin{figure*}
  \centering
\includegraphics[width = \textwidth,trim={0cm 0cm 0cm 0cm},clip]{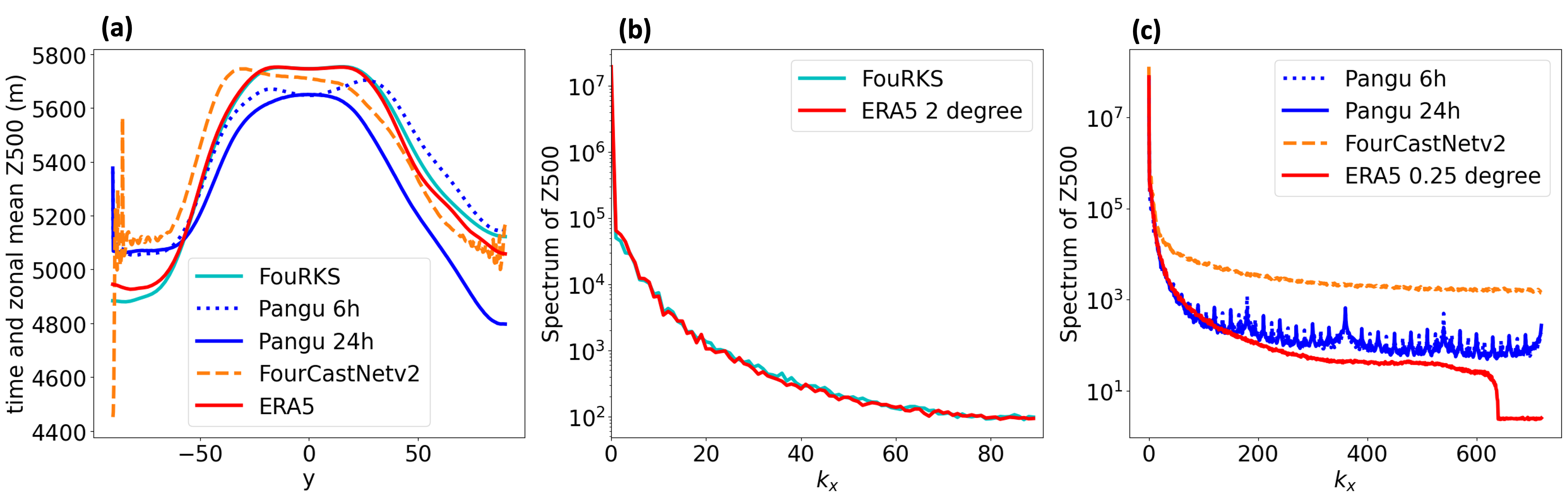}
  \caption{ Long-term statistics of predicted Z500 with FouRKS calculated over $10$ years of autoregressive prediction as compared to the other AI-based weather models. (a) Zonal- and time-mean averaged Z500, predicted by FouRKS and other AI-based weather models as compared to ERA5. (b) Latitude-averaged Fourier spectrum of Z500 computed from FouRKS' after 10 years of prediction as compared to the true ERA5 $2^{\circ}$ data on which FouRKS is trained. (c) The other AI-based weather models do not follow the spectrum of $0.25^{\circ}$ on which they are trained. Both the mean and spectrum are accurately captured by FouRKS while the other models show unphysical characteristics.}
  \label{fig:Z500}
\end{figure*}

\section{Discussion}\label{sec12}
In this paper, we report three major contributions to the field of AI for the weather and climate system, and more broadly, multi-scale, dynamical systems such as turbulent flows: 

(i) For the first time, we explain the cause of instabilities or physical-inconsistencies in deep learning-based models of such systems and attribute it to spectral bias and subsequent error propagation during autoregressive predictions.  

(ii) We propose FouRKS: an architecture-agnostic framework to mitigate spectral bias and perform convergent time integrations with the data-driven models leading to dampened error propagation.

(iii) Finally, we demonstrate long-term stable and physically-consistent emulations with FouRKS for $300000$ days with data from a QG system and $10$ years with ERA5 data.     

This would facilitate accurate sub-seasonal to seasonal probabilistic weather forecasts, extreme weather forecasts, and replace expensive components of Earth system models with efficient data-driven emulators.

The spectral regularizer proposed in the FouRKS framework directly addresses the spectral bias issue during training. However, several other approaches to curbing spectral bias have been proposed in other studies as well that involve better design of activation functions, random feature kernels, etc \citep{rahaman2019spectral,xu2019frequency,hong2022activation,li2019towards}. Future work should focus on building architectural remedies to spectral bias as opposed to a soft constraint as in the case of our proposed spectral regularizer. It should also be kept in mind that for turbulent flow, the effect of spectral bias becomes much more prominent, especially during autoregressive prediction due to the multi-scale nature of turbulence. Hence, building effective mitigation strategies for spectral bias when building data-driven models for the climate system would require one to consider both turbulence physics and fundamental theoretical properties of deep neural networks at the same time.  

The RK4 integrator in FouRKS dampens the error growth during prediction with the data-driven model. Based on earlier work in Krishnapriyan \textit{et al} \citep{krishnapriyan2022learning}, the RK4 integrator is a convergent integration scheme for neural network-based integrators for dynamical systems. However, the fact that RK4-based integrators are convergent for deep neural networks, is still an empirical result. Future studies would benefit from developing rigorous theories, following in the footsteps of numerical analysis to understand a priori convergence properties of neural ordinary differential equations \citep{chen2018neural}, and generally neural network-based integrators of dynamical systems.

The self-supervised spectrum-correction strategy is an intrusive step in FouRKS that increases the computational cost of autoregressive prediction. However, we have found in this study, that it is an essential component for obtaining accurate climate statistics. Several other studies \citep{stachenfeld2021learned,wikner2022stabilizing}, including ours \citep{chattopadhyay2022long}, have found that training with noise, or training with variational models can improve the stability of autoregressive predictions. While there is an absence of rigorous theory to explain the effect of noise on stability, more focused studies need to be directed along these lines to better understand the effect of noise, during training, on autoregressive prediction. In this study, we conducted significant experimentation with different levels of white noise added to the labels during training. It did not improve the stability of the model. The results from those experiments have been reported in Fig. S$3$. It must also be noted that in~\citep{guan2024lucie}, FouRK was sufficient to stabilize the global climate emulator albeit at much lower resolution. The need for the SSL layer depends on the resolution at which the emulation is taking place which determines how quickly small-scale errors grow in time.

While FouRKS allows the data-driven model to produce long-term mean and variability correctly, it is a long way away from being a fully functional climate model. Firstly, data-driven models cannot generalize beyond the current climate, since it has been trained on the current climate and does not have a mechanism to account for radiative forcing. While one can easily fine-tune (transfer learn) data-driven models on future simulations from state-of-the-art climate models, it would then suffer from the same numerical biases as that of climate models. Moreover, there are several other sanity checks that a climate model needs to pass before becoming operational. For example, a large body of literature analyses the responses of certain state variables predicted by climate models to small perturbations of certain other state variables \citep{hassanzadeh2016linear,watson2022climatebench}. For example, a recent study by~\cite{hakim2023dynamical} evaluates Pangu3D in terms of how well it responds to equatorial heating. These tests need to be meticulously performed and evaluated on data-driven models for climate as well.

\section{Data and Methods}\label{sec:data}
\subsection{Reanalyis data and state-of-the-art AI-based weather models}
\label{sec:ERA5}
To evaluate the AI-based weather models, we have considered two different resolutions of ECMWF reanalysis 5 (ERA5) data \citep{hersbach2020era5}. FourCastNetv2 \citep{bonev2023spherical}and GraphCast~\citep{lam2022graphcast} is trained on $0.25^{\circ}$ resolution ERA5 and integrated with a time step of $\Delta t = 6$h, while two different versions of Pangu3D~\citep{bi2022pangu} has been explored, one with $\Delta t = 6$h and the other with $\Delta t = 24$h. FourCastNetv2 uses a spherical Fourier neural operator in its architecture \citep{guibas2021efficient}. Pangu3D is a 3D vision transformer with temporal aggregation and GraphCast is a multi-level graph neural network that has reported short-term skills superior to state-of-the-art numerical weather prediction models such as IFS. Essential aspects of the training of FourCastNetv2 can be found in Bonev \textit{et al.} \citep{bonev2023spherical}, GraphCast in Lam \textit{et al.} \citep{lam2022graphcast}, and  Pangu3D in Bi \textit{et al.}~\citep{bi2022pangu}. The U-NET~\citep{weyn2019can,weyn2020improving,chattopadhyay2021towards} is trained on $2^{\circ}$ resolution ERA5 data and integrated with a time step of $\Delta t = 1$ day. For the U-NET, training is performed on Z500, Z50, and Z850 data between $1979-2015$, validation is performed on the years of $2016$ and $2017$, and testing is performed by initializing the trained model with $12$ independent initial conditions obtained from the year of $2018$. 

\subsection{The two-layer quasi-geostrophic (QG) system} 
\label{sec:QG}
The performance of the data-driven model is also evaluated on the two-layer QG system. The QG system follows the set up as described in Philips~\citep{phillips1951simple} with a barocilinically unstable atmospheric jet. While the QG is not a comprehensive climate model, it is a reasonable representation of the essential aspects of the mid-latitude dynamics of the atmospheric jetstream and is an appropriate testcase for a fully turbulent flow. The full states of the system are given by the upper- and lower-level stream functions, $\psi_1$ and $\psi_2$. The equations solved in the two-layer QG system can be found in Lutsko \textit{et al.}~\citep{lutsko2015applying} and Nabizadeh \textit{et al}.~\citep{nabizadeh2019size}.

For training on QG, we have used the full-state vectors, $\psi_1$ and $\psi_2$ from $8$ independent ensembles, each having $1500$ days of simulated flow. For validation, we have used another independent ensemble having $1500$ days of simulated flow. For testing, $30$ initial conditions from an independent ensemble have been used to intialize the trained model. A time step, $\Delta t = 40 \Delta t_n$, has been used for training and autoregressive predictions, where $\Delta t_n$ is the time step used in the numerical simulation.

\subsection{Baseline U-NET}
\label{sec:Unet}
Here, we describe how the U-NET, as shown in Fig~\ref{fig:fourks_schematic}, is trained to integrate the states of the system (both QG and ERA5) in time. The choice of U-NET is inspired from prior work in data-driven weather forecasting, where U-NET, and variants of U-NET, have been successfully used to predict atmospheric variables in short time scales~\citep{weyn2019can,weyn2020improving, weyn2021sub,chattopadhyay2021towards}. The dynamics of both systems can be described as a PDE, with state vector, $\mathbf{X}$. The following mathematical treatment does not depend on whether we have access to the full states (as in QG) or partial states (as in ERA5) of the system. The evolution of the state, $\mathbf{X}$, is given by:

\begin{align}
\frac{d\mathbf{X}}{dt}= \mathbf{F}\left(\mathbf{X}\left(t\right)\right).
\label{eq:dyn_sys}
\end{align}
The U-NET is trained on pairs of samples of $\mathbf{X}(t)$ and $\mathbf{X}(t+\Delta t)$. For QG, $\mathbf{X}$ consists of $\psi_1$ and $\psi_2$, while for the ERA5 data, it is Z500, Z50, and Z850. 
The objective of the U-NET, $\mathcal{M}\left[\circ,\phi \right]$, where $\phi$ are the trainable parameters, is to predict the right hand side of the equation:

\begin{eqnarray}
   \mathbf{X}(t+\Delta t)=\underbrace{\mathbf{X}(t)+ 
      \int_{t}^{t+\Delta t}{ \mathbf{F}\left(\mathbf{X}\left(t\right)\right)}dt}_{\mathcal{M}\left[\mathbf{X}(t),\phi \right]}.
      \label{eq:direct_pred}
\end{eqnarray}
During inference, $\mathcal{M}$ is used to autoregressively predict the future states of the system, starting from an initial condition, $\mathbf{X}\left(t=0\right)$, obtained from an unseen test set.

The U-NET, $\mathcal{M}$, consists of $6$ hidden convolutional layers without pooling, an input layer, and an output layer. The first $4$ hidden layers have $64$ filters each. The second-last hidden layer has $128$ filters while the last hidden layer has $192$ filters. ReLU activations are used in the input and hidden layers, while the output layer has a linear activation function. A mean-squared error loss function is used. The loss is optimized using stochastic gradient descent with a fixed learning rate of $\alpha =10^{-9}$. The hyperparameters of the U-NET are been obtained after significant trial and error. $\mathcal{M}$ has $\approx 10^5$ trainable parameters.

\subsection{FouRKS: FOUrier-Runge-Kutta-with-Self-supervision}
\label{sec:fourks_method}
In this paper, we propose an architecture-agnostic framework, FouRKS, which employs a three-pronged approach to mitigate instabilities and unphysical features in the data-driven model's predicted flow. The three prongs of FouRKS are:
\begin{itemize}
    \item Fourier-based spectral regularization
    
    \item $4^{th}$-order Runge-Kutta integrator 
    
    \item Self-supervised spectrum correction
\end{itemize}
The first two prongs are applied during training of the model as shown in Fig~\ref{fig:fourks_schematic} in the third row and the third prong is applied during autoregressive prediction with the model as shown in Fig~\ref{fig:fourks_schematic} in the fourth row.

\subsubsection{Fourier-based spectral regularization}
\label{sec:spec_regul}
The first prong aims to penalize the high-wavenumber part of the latitude-averaged absolute value of zonal Fourier coefficients of the predicted fields so as to mitigate the effects of spectral bias. A spectral regularization term, $\mathbf{\mu} \left(\theta\right)$, is used in the loss function, given by :

\begin{align}
  \mu \left(\theta\right)=\sum_{t=0}^{t=T}\biggr\| \widehat{\mathbf{X}}\left(t+\Delta t\right)\biggr \rvert_{{k_x \geq k_T}}-\widehat{\mathbf{H}}\left[\mathcal{N}\left(\mathbf{X}\left(t\right),\theta\right) \right]\biggr \rvert_{{k_x \geq k_T}}\biggr\|_2^2,
\end{align}
such that the total loss function, $\mathbf{L}\left(\theta\right)$, is:

\begin{align}
\label{eq:FourK_spectal_loss}
    \mathbf{L}\left(\theta\right)=\sum_{t=0}^{t=T}\|\mathbf{X}\left(t+\Delta t\right)- \mathbf{H}\left[\mathcal{N}\left(\mathbf{X}\left(t\right),\theta\right) \right]\|_2^2 + \lambda \mu(\theta),
\end{align}
where $\mathcal{N}$ is an U-NET that represents the right-hand-side of Eq.(\ref{eq:dyn_sys}). More details on $\mathcal{N}$ is given in section~\ref{sec:methods_RK4}. $\widehat{\left[\circ\right]}$ represents the absolute zonal Fourier coefficients averaged over all latitudes. However, we have found that, if we compute the latitude-averaged Fourier coefficients only over the midlatitudes for ERA5 and over the region of the jetstream in QG, the performance of FouRK or FouRKS is not affected. $T$ is the total number of temporal samples (across all the ensembles for QG and all the years for ERA5) that are used for training. $\lambda$ is the regularization constant (chosen as $0.8$ in this paper after significant trial and error). $k_x$ is the zonal wavenumber and $k_T$ (chosen as $30$ for QG and $40$ for ERA5 after significant trial and error) is the  threshold wavenumber beyond which the absolute value of Fourier coefficients are penalized in $\mathbf{\mu}\left(\theta\right)$. Here, $\mathbf{H}\left[\circ\right]$ is a $4^{th}$-order Runge-Kutta (RK4) integrator implemented as a differentiable layer inside the architecture. Details on $\mathbf{H}$ is further explained in section~\ref{sec:methods_RK4}.

\subsubsection{A $4^{th}$-order Runge-Kutta (RK4) integrator}
\label{sec:methods_RK4}
The second prong employs a RK4 time-integrator inside the architecture to dampen error propagation. Here, instead of directly predicting $\mathbf{X}(t+\Delta t)$, as done in Eq.~(\ref{eq:direct_pred}), we represent $\mathbf{F}\left(\mathbf{X}\left(t\right)\right)$ in Eq.~(\ref{eq:dyn_sys}) with an U-NET, $\mathcal{N}\left[\circ,\theta\right]$, with trainable parameters, $\theta$:

\begin{eqnarray}
     \mathbf{X}(t+\Delta t)=\underbrace{\mathbf{X}(t)+ 
      \int_{t}^{t+\Delta t} \underbrace{\mathbf{F}\left(\mathbf{X}\left(t\right)\right) dt}_{\mathcal{N[\circ,\theta]}}}_{\mathbf{H[\circ]}}. 
      \label{eq:RK4_pde}
\end{eqnarray}
A custom layer, $\mathbf{H}[\circ]$, performs the integration between $t$ and $t+\Delta t$ via the RK4 scheme following Krishnapriyan \textit{et al.}~\citep{krishnapriyan2022learning} as shown in Eq.~(\ref{eq:RK4_pde}). The operations in the $\mathbf{H}$ are given by:

\begin{subequations}
\begin{align}
        i_1 &= \mathcal{N}\left[\mathbf{X}\left(t\right),\theta \right], \\
        i_2 &= \mathcal{N}\left[\mathbf{X}\left(t\right)+\frac{1}{2}i_1,\theta \right], \\
        i_3 &= \mathcal{N}\left[\mathbf{X}\left(t\right)+\frac{1}{2}i_2,\theta \right], \\
        i_4 &= \mathcal{N}\left[\mathbf{X}\left(t\right)+i_3,\theta \right], \\
        z &= \mathbf{X}\left(t\right)+\frac{1}{6}\left(i_1+2i_2+2i_3+i_4\right).
    \end{align}
\end{subequations}
The predicted state is given by $z= \mathbf{H}\left[\mathcal{N}\left(\mathbf{X}\left(t\right)\right),\theta \right]$.
We denote any architecture equipped with the spectral regularizer and the RK4 integrator as FouRK. The number of parameters in the FouRK framework, where $\mathcal{N}\left[\circ,\theta\right]$ has the same architecture as described in section~\ref{sec:Unet}, is $\approx 10^{5}$. It should be noted that the choice of using $\mathbf{H}$ as an RK4 integrator is because it is convergent when used inside a neural network for integrating dynamical systems \citep{krishnapriyan2022learning}. Other integrators with similar convergence properties can also be used. 

\subsubsection{Self-supervised spectrum correction}
\label{sec:SSL}
This third component is essential in the FouRKS framework to make FouRK stable for long-term climate simulations. FouRKS' third prong is a self-supervised spectrum-correction strategy applied during autoregressive prediction. Here, at every $s$ days ($s = 2$ in both QG and ERA5), FouRK updates its last two layers by implicitly optimizing a new loss function, $\mathbf{L}_1\left(\theta_1\right)$,

\begin{eqnarray}
\label{eq:L1}
        \mathbf{L_1}\left(\theta_1\right)=\sum_{t=0}^{t=T}\biggr\| \left<\widehat{\mathbf{X}}_{train}\right> \biggr \rvert_{k_x \geq k_{T_{S}}} -\widehat{\mathbf{H}} \left[\mathcal{N}\left(\mathbf{X}\left(t\right),\theta\right) \right] \bigg \rvert_{k_x \geq k_{T_{S}}} \biggr \|_2^2,
\end{eqnarray}
where $\theta_1$ are the parameters of the last two layers of $\mathcal{N}$ in FouRK. Here, $\left<\widehat{\mathbf{X}}_{train}\right>$ is the latitude-averaged zonal Fourier spectrum of the state, $\mathbf{X}$, averaged over the training set, where, $\left<\circ\right>$ denotes averaging over training set. $k_{T_{S}}$ is the threshold wavenumber chosen to be $20$ for QG and $30$ for ERA5 after several trials. Here the Fourier spectrum of $\mathbf{X}$ is invariant, i.e. in both the QG and ERA5 data, it does not change in time. This is because the QG system is stationary and while the true weather system is actually non-stationary, the variables that we are emulating (Z50, Z500, Z850) show a relatively constant structure of the Fourier spectrum over time. Thus, optimizing $\mathbf{L}_1\left(\theta_1\right)$ does not require any data. Here, $s$ and $k_{T_{S}}$ are hyperparameters which are obtained after extensive trial and error. 

For this prong to be effective, another U-NET, $\mathcal{N}'\left[\circ,\theta^{'}\right]$, is trained to learn a map from $\widetilde{\mathbf{X}}(t)$ to $\mathbf{X}'(t+\Delta t)$ as shown in Fig.~\ref{fig:fourks_schematic}, fourth row, where $\widetilde{\left[\circ\right]}$ is a sharp spectral filter with cut-off wavenumber as $k_{T_{S'}}$ ($k_{T_{S'}}=30$ for QG and $k_{T_{S'}}=40$ for ERA5 after significant trial and error), and $\mathbf{X}'=\mathbf{X}-\mathbf{\widetilde{X}}$. It uses the same loss function as in Eq.~(\ref{eq:FourK_spectal_loss}) during training, where $\mathbf{X}(t+\Delta t)$ is now replaced by $\mathbf{X}^{'}(t+\Delta t)$. Similar to FouRK, this U-NET also undergoes a spectrum correction every $s$ daysduring autoregressive prediction with loss function, $\mathbf{L}_2\left(\theta^{'}_2\right)$, given by:
\begin{eqnarray}
\label{eq:L2}
        \mathbf{L_2}\left(\theta^{'}_2\right)=\biggr\| \left<\widehat{\mathbf{X'}}_{train}\right> \biggr \rvert_{k_x \geq k_{T_{S'}}} -\widehat{\mathcal{N'}}\left(\mathbf{\widetilde{X}}\left(t\right),\theta^{'}\right)  \bigg \rvert_{k_x \geq k_{T_{S'}}} \biggr \|_2^2.
\end{eqnarray}
  $\theta^{'}_2$ are the parameters of the last two layers of the U-NET, $\mathcal{N}'$. The value of $k_{T_{S'}}$ is a hyperparameter obtained through extensive trial and error. It should be carefully noted that $\mathcal{N'}$ does not evolve a dynamical system. It is a map that learns to predict $\mathbf{X}'$ from $\widetilde{\mathbf{X}}$, where $\widetilde{\mathbf{X}}$ is obtained by filtering the autoregressive prediction of $\mathbf{X}$ from FouRK. Hence, the burden of accurately predicting the state of the system lies on FouRK, while $\mathcal{N'}$ is responsible for predicting the small scales, only one time step ahead.    

At every $s$ days, the output from FouRK whose spectrum is corrected is filtered through a sharp spectral filter with cut-off wavenumber as $k_{T_{S'}}$, and added to to the predicted $\mathbf{X}'(t+\Delta t)$ as shown in Fig.~\ref{fig:fourks_schematic}, to obtain the modified state, $\mathbf{X}^{u}(t+\Delta t)$ at $t+\Delta t$:

\begin{eqnarray}
\label{eq:SSL_update}
\mathbf{X}^{u}(t+\Delta t)=\widetilde{\mathbf{X}}(t+\Delta t)+\mathbf{X}'(t+\Delta t).
\end{eqnarray}

\section*{Acknowledgements}
We thank Alistair Adcroft, Ebrahim Nabizadeh, Karthik Kashinath, Jaideep Pathak, and Laure Zanna for insightful comments and discussions. This work was supported by an award from the ONR Young Investigator Program (N00014-20-1-2722), a grant from the Schmidt Futures Program, and  NASA grant 80NSSC17K0266 to P.H. Computational resources were provided by NSF XSEDE (allocation ATM170020) to use Bridges GPU and the Rice University Center for Research Computing. AC acknowledges National Science Foundation (grant no. 2425667) and computational support from NSF ACCESS MTH240019 and NCAR CISL UCSC0008, and UCSC0009. The codes for FouRKS are publicly available at \url{https://github.com/ashesh6810/FouRKS}.


\bibliographystyle{abbrvnat}

\bibliography{pnas-sample}

\end{document}